# Physically based 3D finite element model of a single mineralized collagen microfibril

Ridha Hambli and Abdelwahed Barkaoui

PRISME laboratory, EA4229, University of Orleans

Polytech' Orléans, 8, Rue Léonard de Vinci 45072 Orléans, France

Phone : +33 (0)2-38-49-40-55

Mail : ridha.hambli@univ-orleans.fr

**ABSTRACT**

Mineralized collagen microfibrils in human bone provide its mechanical properties (stiffness, elasticity, ductility, energy dissipation and strength). However, detailed 3D finite element models describing the mechanical behaviour of the mineralized collagen microfibrils are still lacking. In the current work, we developed a 3D finite element model of the mineralized collagen microfibril that incorporates the physical 3D structural details. The model components consist of five tropocollagen molecules, mineral hydroxyapatite and intermolecular cross-links joining primarily the ends of the tropocollagen molecules. Dimension, arrangement and mechanical behaviour of the constituents are based on previously published experimental and theoretical data. Tensile load was applied to the microfibril under different conditions (hydrated and dehydrated collagen) to investigate the relationship between the structure and the mechanical behaviour of the mineralized collagen microfibril (stress-strain curve and elastic modulus). The computational results match the experimental available data well, and provide insight into the role of the phases and morphology on the microfibril behaviour. Our predicted results show that the mechanical properties of collagen microfibrils arise due to their structure and properties. The proposed 3D finite element model of mineralized collagen microfibril contributes toward the investigation of the bottom-up structure-property relationships in human bone.

**Introduction**

Bone is a hierarchically structured material beginning at a nano-scale level continuing to the whole bone geometry (Fig.1) [1-5]. At the nano-scale level, bone tissue is a composite material composed of an organic phase, consisting mainly of the protein-based material collagen, and a mineral phase, consisting primarily of hydroxyapatite (HA) [2,6].

# Figure 1

At the lowest hierarchical level, bone structure is composed of tropocollagen (TC) which can be viewed as a rod of about 300 nm long and of about 1.1-1.5 nm in diameter, made up of three polypeptide strands, each of which is a left-handed helix [7-9]. The arrangement of the TC molecules comes from the strong chemical bonds (cross-linking) that form between adjacent collagen molecules throughout the collagen bundles [10-13].

During bone remodeling process, TC molecules are deposited first by osteoblasts and arranged with a longitudinal stagger and then thin mineral particles are nucleated to form microfibrils [14, 15]. It is generally believed that mineral particles are nucleated primarily inside the gap region of the microfibrils and fibrils where nucleation sites could be present and where more space is available. In a later stage, however, the mineral particles extend into the extra-collagenous region [16]. The fraction of mineral that is extrafibrillar is not well-established, although Bonar et al. 1985 [17] suggests that as much as 70-75% of the mineral may be extrafibrillar.

At ultrastructure level, mineral and TC molecules are arranged into higher hierarchical levels to form microfibrils, fibrils and fibers [10, 11, 12]. The existence of sub-structures in collagen fibrils have been a debate for years [3, 6-12, 14, 17, 18-27]. Recent studies suggested the presence of microfibrils in fibrils. A longitudinal microfibrillar structure was visualized in both hydrated (wet) [2, 3] and dehydrated (dry) [6]. 3D image reconstructions of collagen fibrils also showed a 4 nm repeat in a transverse section, which was related to the microfibrillar structure [28]. Using X-ray diffraction, Orgel et al. (2006) [24] suggested the presence of right-handed supertwisted microfibrillar structures in collagen fibrils.

A comprehensive study of bone mechanical properties must investigate the tissue at several levels of organization in order to gain a complete understanding of the influence of structure and composition on these properties [1, 5, 11, 29].

Experimental works were performed to investigate the ultrastructure of mineralized collagen. Some authors focused on the mechanical properties of individual collagen fibrils [25, 26, 30]. In addition, mechanical models for mineralized collagen fibrils have been developed by several authors in order to study the effects of the collagen-mineral deformation process [20], to estimate the mechanical properties of mineralized collagen fibrils and bone tissues [22, 31] and to model the 3D orthotropic elastic properties of a single collagen fibril [32]. Nikolov and Raabe (2008) [33] proposed a homogenization method to model the elastic properties of bone at the level of mineralized collagen fibrils from the staggered arrangement of collagen molecules up to an array of parallel mineralized fibrils.

Also, molecular dynamics simulations have been developed to investigate the mechanical properties and the deformation of bone and its constituents at the nanoscale considering the collagen molecules [10, 23, 27, 34-39].

Other experimental studies were performed on hydrated collagen microfibril in the small strain regime based on X-ray diffraction [40] and atomic force microscopy (AFM) [25, 41] to investigate the stress-strain relation and the elastic properties of microfibrils.

However, limited computation works were performed at microfibril level [11]. As a result of these limitations, it is not clear how microfibrils react to mechanical load in either the elastic or plastic deformation regimes. Recently, Gautieri et al. (2011) [11] developed a 2D atomistic collagen microfibril dynamic model that incorporates the full biochemical details of the amino acid sequence of constituting molecules and the nanoscale molecular arrangement to describe the mechanical behaviour at the microfibril level. Unfortunately, due to the size of the model, full microfibril model simulations require enormous computational power and are still not affordable at present. However, 3D FE models of a single microfibril structure considering the TC, mineral phases and the cross-links are still lacking.

In this study, we expand upon previous models dealing with bone ultrastructure modeling in three aspects: (i) Creation of lower level (microfibril) realistic 3D finite element (FE) model to represent the structure of the mineralized collagen microfibril with three constituents (mineral, TC molecules and cross-links). (ii) Investigation of the microfibril deformation behaviour in relation with its constituents states (hydrated, dehydrated collagen, mineral filling the gap or extra-collagenous region). (iii) Validation of the model based on experimental data from literature (stress-strain curve and elastic modulus).

The proposed 3D FE model of mineralized collagen microfibril enables the bottom-up investigation of structure-property relationships in human bone. Such model can be applied to study the effects of biochemical details related to the collagen or the mineral components on the strength of human bone.

## 2. Method

In the current section, We will discuss (i) the model development (on the basis of a descriptive experimental data) and (ii) the presentation of the behaviour law of each phase (on the basis of mechanical experimental data).

Model validation will be discussed in results and discussion section. The predicted microfibril elastic modulus and the stress-strain curve will be compared with available experimental data.

### 2.1. Microfibril Model Composition

Smith (1968) [42] suggested that each microfibril consists of exactly five molecules in a generic circular cross section. Lee et al. (1996) [43] suggested that collagen microfibrils have a quasi-hexagonal structure. In the current work, the smith's model (cylindrical microfibril) was retained for simplicity to develop the 3D microfibril FE model. In addition, the current work investigate the microfibril mechanical behaviour under tensile load which depends mainly on the area of the cross section (section shape has no influence under tensile load).
The elementary components of the microfibril can be distinguished as follow:

(a) **Tropocollagen**: Long cylindrically shaped TC molecules establishing a continuum in which mineral crystals are embedded. The mineralization process reduces the lateral spacing between collagen molecules and bringing them closer to each other [18, 44]. It have been reported by Lees (1981) and Fratzl et al. (1993) [7,8] that effective molecular diameter of dry TC molecules is 1.09 nm and that of wet TC molecules is about 1.42-1.5 nm. Hydration affects the packing (molecules are closer) and the sliding between molecules (which become more difficult due to increased adhesion). TC length is about 300 nm [10, 11] self-assembled in the form of microfibrils.

(b) **Mineral**: Plate or needle-shaped mineral crystals consisting of impure HA ($Ca_{10}[PO_4]_6[OH]_2$) with typical 1–5 nm thickness, and 25–50 nm length [45]. This longest dimension is typically found to be oriented parallel to the collagen molecules.
During bone remodeling process, mineral particles are nucleated primarily inside the gap region of the microfibrils. In a later stage, the mineral particles extend into the extra-collagenous region [16].

(c) **Cross-links**: The microfibril structure is stabilized through intermolecular cross-links joining two TC molecules. Cross-linking of tropocollagen molecules play a critical role in bone microfibrils, fibrils and fibers connectivity [21, 44, 46-48]. Cross-linking is either enzymatically or non-enzymatically mediated [49, 50]. It have been reported that the formation of intermolecular covalent cross-links have a significant effects on material properties (strength and brittleness) [51-58] and mechanical behaviour [50, 59-61]. Considering a macroscopic response of bone, enzymatic cross-linking has been linked to improved mechanical properties [62] whereas non-enzymatic cross-linking prevents energy absorption by microdamage formation and may accelerate brittle fracture [59, 63-66].

(d) **Different non-collagenous organic molecules**: Predominantly lipids and proteins which regulates HA mineralization, probably by proteins supporting or inhibiting mineralization, possibly also by lipids [67-69].

(e) **Water**: Provides the liquid environment for the biochemical activity of the non-collagenous organic matter.

The mineralized periodic microfibril is a helical assembly of five TC molecules (rotational symmetry of order 5), which are offset one another with apparent periodicity of 67 nm (Fig. 2). This period of dimension length is denoted by the letter D and is used as a primary reference scale in describing structural levels. The helical length of a collagen molecule is 4.47 D ≈ 300 nm and the discrete gaps (hole zones) are about 0.66 D ≈ 44 nm (some references give it as 35 nm) between two consecutive TC molecules. Those five molecules create a cylindrical formation with a diameter 3.5–4 nm and its length is unknown [10, 11].

# Figure 2

In Tab.1 are reported the dimensions of the different constituents shown in Fig.2.

# Table 1

**2.2. Microfibril FE Model Generation**

The proposed 3D FE model of microfibril with symmetric and periodic repetitive portion (one period of length 340 nm of Fig. 2) was generated based on following hypothesis as described in section (**Microfibril Model Composition**):

(a) Tropocollagen molecules are approximated by cylinders with around 300 nm length and 1.09 (dry) and 1.5 nm (wet) diameters.
(b) Mineral phase is modeled as homogenized material with uniform concentration. Three models are considered here:
   **Model TC**: Microfibril composed of pure TC (without mineral) linked with cross-links.
   **Model GM** (First stage mineral formation): Microfibril composed with TC, mineral filling the gap region of the TC and cross-links.
   **Model EM** (second stage mineral formation): Microfibril composed with TC, mineral filling the extra-collagenous space and cross-links.
(c) Cross links are modeled using calibrated spring elements: Siegmund et al. (2008) [59] developed a 2D FE model to investigate the failure of mineralized collagen fibrils considering the role of collagen cross-linking. The authors suggested that the enzymatic cross-links are placed between the TC molecules ends and non-enzymatic cross-links appear to possess no specific spatial arrangement with a random distribution between collagen–collagen domains. In current work, following Siegmund et al. (2008)[59] cross-links representation, enzymatic cross-links were modeled by spring elements joining two TC end terminals and non-enzymatic cross-links are modeled by springs elements joining the TC–TC interfaces at their end surfaces (Fig. 3). The springs rheological behaviour were calibrated with experimental and computed data reported in [11, 59]. Adhesion forces between mineral and TC interfaces remain weak enough to allow for slip in the small strain regime. Therefore, sliding is neglected in the present work.

# Figure 3

(d-e) It have been suggested that non-collagenous organic molecules and water in tissue biomineralisation may be absorbed to crystal surfaces, bound to collagen, or situated freely in the liquid environment between mineral and collagen [67-69]. Therefore, their effects are neglected in the present work for simplicity.

The proposed 3D FE model of a single microfibril composed of previously described components (a-e) is given in Fig. 4. The model is composed of about 22000 tetraedric elements and the simulations were performed using the software ABAQUS/Standard with a computation time of about 45 minute for a complete simulation of a single microfibril model.

The bottom surface of the microfibril was encastred and an uniaxial tensile force ($F$) along the axis of the collagen molecules was applied to the top surface of the microfibril.

In general, bone organ is loaded in compression. Nevertheless, due to eccentric loading convex side of bone undergoes local tension and the concave side of bone undergoes local compression. We present here results for tensile load instead of compressive load on the model for consistency with available experimental data (stress-strain curves under tensile tests and Young modulus measured under tensile tests). Nevertheless, our FE model can be applied to investigate the compressive behaviour of mineralized microfibrils.

From the simulation results, the overall microfibril elongation $\Delta l$ and local strains/stresses fields of the microfibril and its phases were calculated.

Geometry of every constituent and their arrangement are presented (Fig. 4) separately for clarity. Due to the excessive microfibril length to diameter ratio (340/4), only a segment of microfibril full length is plotted in Fig. 4.

# Figure 4

Current study focuses on the FE description and the elastic behaviour of the mineralized collagen microfibril in the small strain regime ($\varepsilon < 5\%$) [11]. Plasticity, rupture and relative sliding on the interface between TC and mineral were not considered here.

Stress-strain curves were predicted with the FE model and the corresponding elastic modulus of the microfibril was compared to previously published experimental data [25, 40, 41] and theoretical analyses [11].

An alternative method to calculate the apparent tensile stress $\sigma_{app}$ and the corresponding apparent strain $\varepsilon_{app}$ applied to the microfibril can be calculated using:

$$\sigma_{app} = \frac{F}{A} \qquad (1)$$

and

$$\varepsilon_{app} = \frac{\Delta l}{l_0} \qquad (2)$$

where $A$, $\Delta l$ and $l_0$ are respectively the apparent area of the microfibril, the total microfibril elongation computed by the FE analysis and the initial length of the microfibril.

The apparent elastic modulus of the microfibril model is then defined as the apparent applied stress divided by the apparent strain of the microfibril by:

$$E_{app} = \frac{\sigma_{app}}{\varepsilon_{app}} \qquad (3)$$

### 2.3. Tropocollagen mechanical behaviour

Experimental studies dealing with collagen mechanical behaviour [10, 36] and mechanical properties [71-73] of single TC molecule have been investigated by several authors.
Sasaki and Odajima (1996b) [74] studied the stress-strain behaviour of single collagen molecule. Using X-ray diffraction and simultaneous tensile loads on molecules, a linear stress-strain behaviour was deduced for collagen molecules in the hydrated state. The deduced Young's modulus of collagen was ranged from 2.8 GPa to 3.0 GPa.

### 2.4. Mineral mechanical behaviour

The mineral in bone is HA, which is a calcium phosphate $Ca_{10}(PO_4)_6(OH)_2$. It have been reported that HA mineral is stiff and extremely fragile exhibiting elastic isotropic behaviour [16, 59, 66, 75]. Its experimental Young's modulus is about $E_m = 114\,GPa$ and Poisson's ratio is about $v_m = 0.3$. The mineral fails in brittle mode at a stress of $\sigma_m^u \approx 100\,MPa$. Usually the two phases (mineral in gap and mineral in extra-collagenous regions) are modeled as linear elastic [16].

The mechanical properties of the TC and the mineral reported in the literature are summarized in Table 2.

# Table 2

### 2.5. Cross-links mechanical behaviour

From a FE point of view, spring elements can be used to model connections between two different regions (TC ends here) to couple a force with a relative displacement representing the elasticity of the physical connecting constituents (physical cross links). The spring model retained in the current work acts between two nodes belonging to two different cross-linked TC molecules. The line of action corresponds to the line joining the two nodes, so that this line of action can rotate in large-displacement analysis.

In present work, in order to describe the FE behaviour of a cross-link as a constitutive model, we incorporate a non-linear spring elements with three calibrated rheological regimes as suggested recently by [12] (Fig. 5): (i) Elastic regime with (ii) delayed response due to the unraveling of the telopeptide and (iii) the friction representing the intermolecular slippage which can be modeled by a perfectly plastic spring response.

# Figure 5

Rheological spring parameters were calibrated based on studies referenced in Tab. 3.

# Table 3

### 3. Results and discussion

In the current work, the analyses focused on simulation of the whole microfibril deformation and the prediction of the apparent stress-strain curve and corresponding apparent elastic modulus. Accordingly, the insights offered by the analyses mainly focus on overall mechanistic understanding of the microfibril mechanical properties variation providing variation on the phases contents and the properties of each constituent.

Fig. 6 shows the equivalent von Mises stress distribution within the microfibril for three different models (a) **TC model**, (b) **GM model** and (c) **EM model**. Note that the TC molecules are cross-linked for the three models. The contours were plotted for a fraction of the microfibril at a longitudinal cross section for clarity.

# Figure 6

Observations of the different contours reveal different stress distribution within the microfibril components due to the structural difference of the mineral phase ratio and the nature of the TC (dry or wet). Our predicted results show that adding mineral to the pure collagen microfibril leads to increase of the distribution of equivalent stress revealing increase of the microfibril stiffness in combination with the TC molecules.

Burstein et al. (1975) [77] studied the mechanical properties of compact bovine. They showed that the Young's modulus increases with the degree of mineralization. Bowman et al. (1996) [78] investigated the tensile behaviour of completely demineralized bovine cortical bone and reported that average tensile Young's modulus, was much lower than untreated compact bone. They noted an initial non-linear 'toe' region at small strain regime <4% with low elastic modulus value which was attributed to uncoiling of the collagen molecules. At higher strains, the behaviour was linear with higher elastic modulus attributed to stretching of the collagen molecules.

Fig. 7 plots the stress–strain responses for TC, GM and EM models, under tensile loading. The averaged equivalent strain ($\varepsilon$) was computed for every phase (*i*) based on the relation:

$$\varepsilon^i = \frac{1}{V_o} \int_o \varepsilon_j \, dV \qquad (4)$$

Where $V_o$ and $\varepsilon_j$ denote respectively the phase reference domain and the output at every finite element location *j*.

For tensile strains in the elastic regime up to 4%, the stress–strain responses for the models (TC, GM and EM) exhibit different slopes, indicating that increase of mineral crystals during bone formation significantly increases the bone stiffness. Mechanical fundamental function of bone is to provide a lightweight frame to support the body weight. The results reveals how composite bone combines the high stiffness and fragility of the mineral phase and the high toughness and ductility of the collagen phase to fulfill this essential function.

# Figure 7

Very good agreement is obtained between predicted and experimental results in the small strain regime based on X-ray diffraction [40], atomic force microscopy (AFM) [25, 41] and molecular dynamics (MD) computation [11].

Predicted apparent elastic modulus of microfibril variation for the different models are plotted in Fig. 8 for wet and dry states of TC.

# Figure 8

The elastic modulus is strongly influenced by TC states (wet/dry) (Fig. 8-a). Our prediction shows that hydrated TC model experience a more soft-like behaviour compared to dehydrated TC in conformity with finding of [29].

The results reveal that increasing of the mineral phase contents within the microfibril combined with the TC state (dry or wet) leads to increase of the apparent modulus of the microfibril (Fig. 8-b). A direct comparison of the elastic modulus of pure collagen microfibril model versus that of mineralized collagen microfibrils suggests that the microfibril stiffness is highly dependent of the mineral content (gap or extra-collagenous) (Fig. 8-b). The results suggests that at this ultrastructure scale, the elasticity is strongly scale dependent. This finding agrees well with experimental and numerical results reported by [11].

In addition, the results show that change of the elastic modulus when comparing a single TC molecule (2.8 GPA) to the predicted pure collagen microfibril model (0.1 to 0.25 GPa) suggest that elasticity is strongly scale dependent. Microfibril atomistic model developed by [11] found that a direct numerical comparison suggests a factor of 10-20 difference in the Young's moduli of single molecule collagen and collagen microfibrils.

In Fig. 9 is plotted the variation of mineral and TC molecules strains obtained for hydrated and dehydrated TC molecules versus the apparent microfibril strain (case of mineral filling the gap region: GM model).

# Figure 9

One can observe that in small strain regime, both mineral strain ($\varepsilon_m$) and TC strain ($\varepsilon_{TC}$) are linearly correlated to the microfibril strain ($\varepsilon_{mf}$). The slope of the curves depend whether the TC is hydrated or not. Also, at a given microfibril strain, the strain sustained by mineral is lower than that of TC due to its elastic modulus.

The predicted results reveals that in the small strain elastic regime, the response results is constant TC-to-microfibril strain ratio ($\varepsilon_{TC}/\varepsilon_{mf}$) of about 0.49 (dehydrated TC) and of about 0.33 (hydrated TC) as well as a constant mineral-to-microfibril strain ratio ($\varepsilon_m/\varepsilon_{mf}$) of about 0.14 (hydrated TC) and (dehydrated TC) of about 0.22

FE results in conformity with experimental ones of [29] considering the cooperative deformation of mineral and collagen in bone at the nanoscale, show that tensile load on microfibril induces a coordinated deformation process at its constituents level. These finding suggest that the hierarchical structure of the microfibril leads to a gradation of the tensile strain applied to the whole microfibril to its constituents (stiff mineral, soft TC and elastic cross-links). Additionally, different ratio values of $\varepsilon_m/\varepsilon_{mf}$ and $\varepsilon_{TC}/\varepsilon_{mf}$ suggest that load transfer between the mineral and the collagen occur by shear transfer (Fig. 10) in the microfibril and involves a load transfer at the mineral-TC interface mediated by the cross-links. Our results are in conformity with Ji (2008) [79] observation. The author has shown that failure of TC–mineral assemblies is a result of combined tensile and shear loading. A stress concentration in the mineral and collagen phases near the collagen/mineral interfaces (Fig. 10) is clearly observed, which may be critical for the properties of the composite and which could induce debonding on the interfaces at the plastic regime.

# Figure 10

At macroscopic level, bone mechanical properties depend on the properties of its elementary constituents (microfibrils composed of TC molecules, mineral and cross-links). The development of 3D FE models to investigate the structure-properties relations for mineralized collagen microfibrils is of crucial importance, for the evaluation of the mechanical properties of bone tissue. In this paper, a 3D FE model was achieved to study the nanomechanical behaviour of collagen microfibril considering the geometry and arrangement of the model constituents and the state of the TC molecules (dry/wet).

The developed 3D FE model to simulate microfibrils behaviour under tensile load has shown that its constituents lead to synergical deformation mechanisms which appear to modulate the apparent behaviour and elastic modulus of the microfibril. Results of the FE model deformation show that the different components of the microfibril take up successively lower levels of strain. The results suggested that the TC can be viewed as a staggered model of load transfer in bone matrix, exemplifying the hierarchical nature of bone deformation to increase the bone ductility in order to protect against its brittle fracture.

The current approach is, to our knowledge, the first 3D FE model describing the mechanical behaviour of microfibril. However, despite the success of the proposed FE model, it will be useful to address the limitations and challenges of this investigation in the future. Specifically, it is important to extend the model to include elastic large strain regime, plasticity and rupture in both phases (TC and mineral) and relative sliding on the interface between two phases. However, the effect of cross-links between molecules and intermolecular sliding has been observed to occur at larger deformation regime [12]. Despite these limitations, the proposed computational model was able to capture the mechanical behaviour observed in experiments. The associated FE prediction can be applied in order to investigate the relationship between bone strength and the normal and pathological mechanical behaviors of collagenous bone tissue.

## Acknowledgements

This work has been supported by French National Research Agency (ANR) through TecSan program (Project MoDos, n°ANR-09-TECS-018).

## References

[1] Rhao JY, Kuhn-Spearing L, Zioupos P. (1998) Mechanical properties and hierarchical structure of bone, medical Engineering &Physics 20, 92-102.
[2] Raspanti M, Congiu T, Guizzardi S (2001) Tapping-mode atomic force microscopy in fluid of hydrated extracellular matrix. Matrix Biol., 20: 601-604.
[3]Habelitz S, Balooch M, Marshall SJ, Balooch G, Marshall GW (2002) In situ force microscopy of partially demineralized human dentin collagen fibrils. J. Struct. Biol., 138: 227-236.
[4] Hambli R, Katerchi H, Benhamou CL (2010) Multiscale methodology for bone remodelling simulation using coupled finite element and neural network computation,Biomechanics and Modeling in Mechanobiology: 10-1, 133-145.
[5] Hambli R (2011) Apparent damage accumulation in cancellous bone using neural networks. J. of the Mech. Behaviour of Biomedical Materials, doi:10.1016/j.jmbbm.2011.03.002.
[6] Baselt DR, Revel JP, Baldschwieler JD (1993) Subfibrillar structure of type I collagen observed by atomic force microscopy. Biophys. J., 65:2644-2655.
[7] Lees S (1981) A mixed packing model for bone collagen. Calcif Tissue Int.,33(6):591-602.
[8] Fratzl P, Fratzl-Zelman N, Klaushofer K (1993) Collagen Packing and Mineralization. An X-Ray Scattering Investigation of Turkey Leg Tendon. Biophys. J., 64:260–266.
[9] Landis WJ, Hodgens KJ, Arena J, Song MJ, McEwen BF (1996) Structural relations between collagen and mineral in bone by high voltage electron microscopic tomography. Microscopy Research and Technique, 33(2): 192-202.
[10] Buehler MJ (2008) Nanomechanics of collagen fibrils under varying cross-link densities: atomistic and continuum studies. Journal of the Mechanical Behaviour of Biomedical Materials 1 (1), 59–67.
[11] Gautieri A, Vesentini S, Redaelli A, Buehler MJ (2011) Hierarchical structure and nanomechanics of collagen microfibrils from the atomistic scale up, Nano Letters 9,11(2):757-66.
[12] Uzel SGM, Buehler MJ (2011) Molecular structure, mechanical behaviour and failure mechanism of the C-terminal cross-link domain in type I collagen. Journal of the Mechanical Behaviour of Biomedical Materials, 4, 153–161.
[13] Christiansen DL, Huang EK, Silver FH (2000) Assembly of type I collagen: fusion of fibril subunits and the influence of fibril diameter on mechanical properties. Matrix Biol 19 409–420.

[14] Glimcher MJ (1987) The nature of the mineral component of bone and the mechanism of calcification. Instr. Course Lect. 36:49–69.

[15] Posner AS (1987) Bone mineral and the mineralization process. In Bone and Mineral Research/5, W. A. Peck, editor. Elsevier Science Publ. B.V., New York, Amsterdam, Tokyo. 65–116.

[16] Pidaparti RM, Chandran A, Takano Y, Turner CH (1996) Bone mineral lies mainly outside collagen fibrils: predictions of a composite model for osteonal bone. J Biomech 29:909–916.

[17] Bonar LC, Lees S, Mook H.A. (1985) Neutron diffraction studies of collagen in fully mineralized bone. J. Mol. Biol. 181:265–270.

[18] Lees S (1987) Considerations Regarding the structure of the Mammalian Mineralized Osteoid From Viewpoint of the Generalized Packing Model. Connect. Tissue Res., 16, 281–303.

[19] Landis WJ, Librizzi JJ, Dunn MG, Silver FH (1995) A study of the relationship between mineral content and mechanical properties of turkey gastrocnemius tendon. J. Bone Mineral Res. 10:859–867.

[20] Jäger I, Fratzl P (2000) Mineralized collagen fibrils: a mechanical model with a staggered arrangement of mineral particles. Biophys. J. 79: 1737–1746.

[21] Gelse K, Poschl E, Aigner T (2003) Collagens—structure, function, and biosynthesis. Advanced Drug Delivery Reviews 55 (12), 1531–1546.

[22] Akkus O (2005) Elastic deformation of mineralized collagen fibrils: an equivalent inclusion based composite model. Trans. ASME. 127:383– 390.

[23] Bozec L, Horton M (2005)  Topography and mechanical properties of single molecules of type I collagen using atomic force microscopy. Biophysical Journal, 88(6): p. 4223-4231.

[24] Orgel J, Irving TC, Miller A, Wess TJ (2006) Microfibrillar structure of type I collagen in situ. Proc. Natl. Acad. Sci. USA, 103: 9001-9005.

[25] van der Rijt JAJ, van der Werf KO, Bennink ML, Dijkstra PJ, Feijen J (2006) Micromechanical testing of individual collagen fibrils. Macromolecular Bioscience, 6(9):697-702.

[26] Eppell SJ, Smith BN, Kahn H, Ballarini R (2006) Nano measurements with micro-devices: mechanical properties of hydrated collagen fibrils. Journal Of The Royal Society Interface, 3(6):117-121.

[27] Bhowmik R, Katti KS Katti DR (2007) Mechanics of molecular collagen is influenced by hydroxyapatite in natural bone. J. Mater. Sci. 42:8795–8803.

[28] Holmes DF, Gilpin CJ, Baldock C, Ziese U, Koster AJ, Kadler KE (2001) Corneal collagen fibril structure in three dimensions: structural insights into fibril assembly, mechanical properties, and tissue organization. Proc. Natl. Acad. Sci. USA, 98:7307-7312.

[29] Gupta HS, Seto J, Wagermaier W, Zaslansky P, Boesecke P, Fratzl P (2006) Cooperative deformation of mineral and collagen in bone at the nanoscale. PNAS, 21-47, 17741–17746

[30] Shen ZL, Dodge MR, Kahn H, Ballarini R, Eppell SJ (2008) Stress-strain experiments on individual collagen fibrils. Biophysical Journal, 95(8): p. 3956-3963.

[31] Fritsch A, Hellmich C (2007) Universal microstructural patterns in cortical and trabecular, extracellular and extravascular bone materials: micromechanics-based prediction of anisotropic elasticity,  J. Theor. Biol. 244:597–620.

[32] Akiva U, Wagner HD, Weiner S (1998) Modeling the threedimensional elastic constants of parallel-fibered and lamellar bone. J. Mater. Sci. 33:1497–1509.

[33] Nikolov S, Raabe D (2008) Hierarchical Modeling of the Elastic Properties of Bone at Submicron Scales: The Role of Extrafibrillar Mineralization. Biophysical Journal (94) 4220–4232.

[34] Cusack S, Miller A (1979) Determination of The Elastic-Constants of Collagen By Brillouin Light-Scattering. Journal of Molecular Biology, 135(1): p. 39-51.
[35] Sun Y, Luo Z, Fertala A, An K (2002) Direct quantification of the flexibility of type I collagen monomer, Biochemical and Biophysical Res. Com., 295(2): p. 382-386.
[36] Buehler MJ, Wong SY (2007) Entropic elasticity controls nanomechanics of single tropocollagen molecules. Biophysical Journal, 93(1):37-43.
[37] Gautieri A, Buehler MJ, Redaelli A (2009) Deformation rate controls elasticity and unfolding pathway of single tropocollagen molecules. J Mech Behav Biomed 2:130-137
[38] Broedling NC, Hartmaier A, Buehler MJ, Gao H (2008) The strength limit in a bio-inspired metallic nanocomposite. J. Mech. Phys. Solids. 56-3, 1086-1104.
[39] Buehler MJ (2006) Atomistic and continuum modeling of mechanical properties of collagen: elasticity, fracture and self assembly. J Mater Res, 21(8):1947–61.
[40] Sasaki N, Odajima S. (1996), Elongation mechanism of collagen fibrils and force-strain relations of tendon at each level of structural hierarchy, J. Biomech. 29 (9), 1131–1136.
[41] Aladin DM, Cheung KM, Ngan AH, Chan D, Leung V., Lim CT, Luk KD, Lu WW (2010) Nanostructure of collagen fibrils in human nucleus pulposus and its correlation with macroscale tissue mechanics. J. Orthop. Res. 28 (4): 497–502.
[42] Smith JW (1968) Molecular Pattern in Native Collagen, Nature, 219:157-158.
[43] Lee J, Scheraga HA, Rackovsky S (1996) Computational study of packing a collagen-like molecule: quasi-hexagonal vs "Smith" collagen microfibril model. Biopolymers. 40(6) 595-607.
[44] Fratzl P, Weinkamer R (2007) Nature's hierarchical materials. Progress in Materials Science 52:1263–1334.
[45] Weiner S, Wagner HD (1998) The material bone. Structure–mechanical function relations. Annu. Rev. Material Sci. 28: 271–298.
[46] Knott L, Bailey AJ (1998) Collagen cross-links in mineralizing tissues: a review of their chemistry, function, and clinical relevance. Bone 22 (3), 181–187.
[47] Eyre DR, Weis MA, Wu JJ (2008) Advances in collagen crosslink analysis. Methods 45 (1): 65–74.
[48] Gupta HS, Seto J, Krauss S, Boesecke P, Screen HRC (2010) In situ multi-level analysis of viscoelastic deformation mechanisms in tendon collagen. Journal of Structural Biology 169 (2), 183–191.
[49] Eyre DR, Dickson IR, Van Ness K (1988) Collagen cross-linking in human bone and articular cartilage. Age-related changes in the content of mature hydroxypyridinium residues. Biochemical Journal 252:495–500.
[50] Bailey AJ (2001) Molecular mechanisms of ageing in connective tissues. Mechanisms of Ageing and Development 122 (7):735–755.
[51] Vashishth D, Wu P, Gibson G (2004) Age-related loss in bone toughness is explained by non-enzymatic glycation of collagen. Transactions of the Orthopaedic Research Society 29.
[52] Wu P, Koharski C, Nonnenmann H, Vashishth D (2003) Loading on non-enzymatically glycated and damaged bone results in an instantaneous fracture. Transactions of the Orthopaedic Research Society 28, 404.
[53] Catanese J, Bank R, Tekoppele J, Keaveny T (1999) Increased crosslinking by non-enzymatic glycation reduces the ductility of bone and bone collagen. In: Proceedings of the American Society for Mechanical Engineering Bioengineering Conference. 42:267–268.
[54] Boxberger J, Vashishth D (2004) Nonenzymatic glycation affects bone fracture by modifying creep and inelastic properties of collagen. Transactions of the Orthopaedic Research Society 29, 0491.

[55] Tang S, Bank R, Tekoppele J, Keaveny T (2005) Nonenzymatic glycation causes loss of toughening mechanisms in human cancellous bone. Transactions of the Orthopaedic Research Society 30.

[56] Wang X, Shen X, Li X, Agarwal CM (2002) Age-related changes in the collagen network and toughness of bone. Bone 31:1–7.

[57] Viguet-Carrin S, Roux JP, Arlot ME, Merabet Z, Leeming DJ, Byrjalsen I, Delmas PD, Bouxsein ML (2006) Contribution of the advanced glycation end product pentosidine and of maturation of type I collagen to compressive biomechanical properties of human lumbar vertebrae. Bone 39:1073–1079.

[58] Allen MR, Gineyts E, Leeming DJ, Burr DB, Delmas PD (2007) Bisphosphonates alter trabecular bone collagen cross-linking and isomerization in beagle dog vertebra. Osteoporosis International, 19 (3):329-337.

[59] Siegmund T, Allen MR and Burr DB (2008) Failure of mineralized collagen fibrils: Modeling the role of collagen cross-linking, Journal of Biomechanics 41, 1427–1435.

[60] Saito M, Marumo K (2009) Collagen cross-links as a determinant of bone quality: a possible explanation for bone fragility in aging, osteoporosis, and diabetes mellitus. Osteoporosis International 21 (2):195–214.

[61] Barkaoui A, Bettamer A, Hambli R (2011) Failure of mineralized collagen microfibrils using finite element simulation coupled to mechanical quasi-brittle damage, Procedia Engineering 10:3185–3190.

[62] Banse X, Sims TJ, Bailey AJ (2002) Mechanical properties of adult vertebral cancellous bone: correlation with collagen intermolecular cross-links. Journal of Bone and Mineral Research 17:1621–1628.

[63] Vashishth D, Gibson GJ, Khoury JI, Schaffler MB, Mimura J, Fyhrie DP (2001) Influence of non-enzymatic glycation on biomechanical properties of cortical bone. Bone 28:195–201.

[64] Tang S, Zeenath U, Vashishth D (2007) Effects of non-enzymatic glycation on cancellous bone fragility. Bone 40:1144–1151.

[65] Nyman JS, Roy A, Tyler JH, Acuna RL, Gayle HJ, Wang X (2007) Age-related factors affecting the postyield energy dissipation of human cortical bone. Journal of Orthopaedic Research 25:646–655.

[66] Vashishth D (2007) The role of collagen matrix in skeletal fragility. Current Osteoporosis Reports 5, 62–66.

[67] Urist M.R., DeLange, R.J. and Finerman G.A.M., 1983. Bone cell differentiation and growth factors. Science 220, 680–686.

[68] Arsenault AL (1991) Image analysis of collagen-associated mineral distribution in crytogenically prepared turkey leg tendons. Calcified Tissue Int 48, 56–62.

[69] Hunter GK, Hauschka PV, Poole AR, Rosenberg LC, Goldberg H.A., 1996. Nucleation and inhibition of hydroxyapatite formation by mineralized tissue proteins. Biochem J 317, 59–64.

[70] Orgel J, Miller A, Irving TC, Fischetti RF, Hammersley AP, Wess TJ (2001) The in situ supermolecular structure of type I collagen. Structure 9 (11), 1061–1069.

[71] An KN, Sun YL, Luo ZP (2004) Flexibility of type I collagen and mechanical property of connective tissue. Biorheology, 41(3–4):239–46.

[72] Sun YL, Luo ZP, An KN (2001) Stretching short biopolymers using optical tweezers. Biochem Biophys Res Commun, 286(4):826–30.

[73] Sun YL, Luo ZP, Fertala A, An KN (2004) Stretching type II collagen with optical tweezers. J Biomech, 37(11):1665–1669.

[74] Sasaki N, Odajima S. (1996b) Stress-strain curve and Young's modulus of a collagen molecule as determined by the X-ray diffraction technique. J. Biomech., 29: 655-658.

[75] Katz J, Ukraincik K (1971) On the anisotropic elastic properties of hyroxyapatite. Journal of Biomechanics. 4, 221–227.

[76] Buehler MJ, Keten S, Ackbarow T (2008) Theoretical and computational hierarchical nanomechanics of protein materials: Deformation and fracture, Progress in Materials Science 53:1101–1241.

[77] Bowman SM, Zeind J, Gibson LJ, Hayes WC, McMahon TA (1996) The tensile behaviour of demineralized bovine cortical bone. J. Biomech. 26:1497–1501

[78] Burstein AH, Zika J, Heiple K, Klein L (1975) Contribution of collagen and mineral to the elastic–plastic properties of bone., J. Bone Joint Surg. A 57:956–961.

[79] Ji BH (2008) A study of the interface strength between protein and mineral in biological materials. J Biomech. 41(2):259–66.

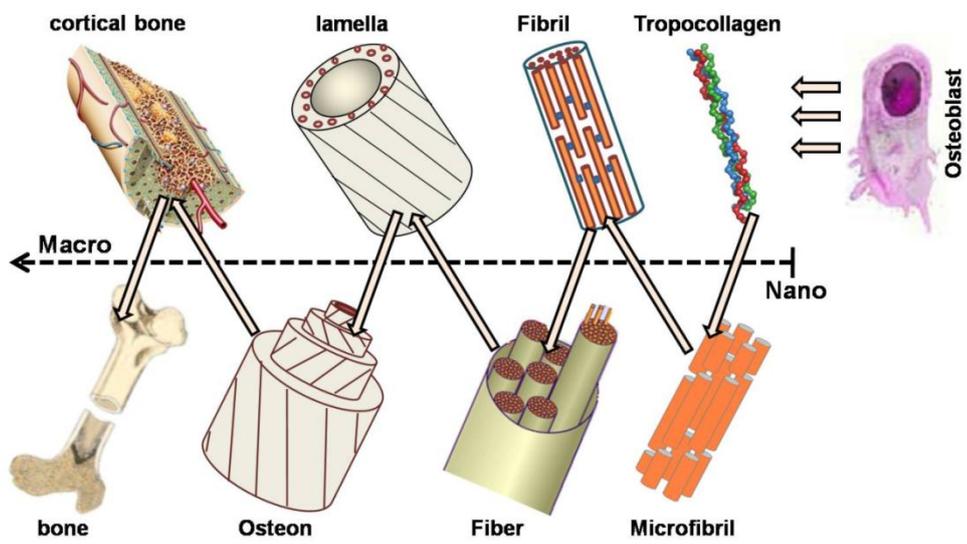

Figure 1. The multiscale hierarchical structure of cortical bone composed of seven levels.

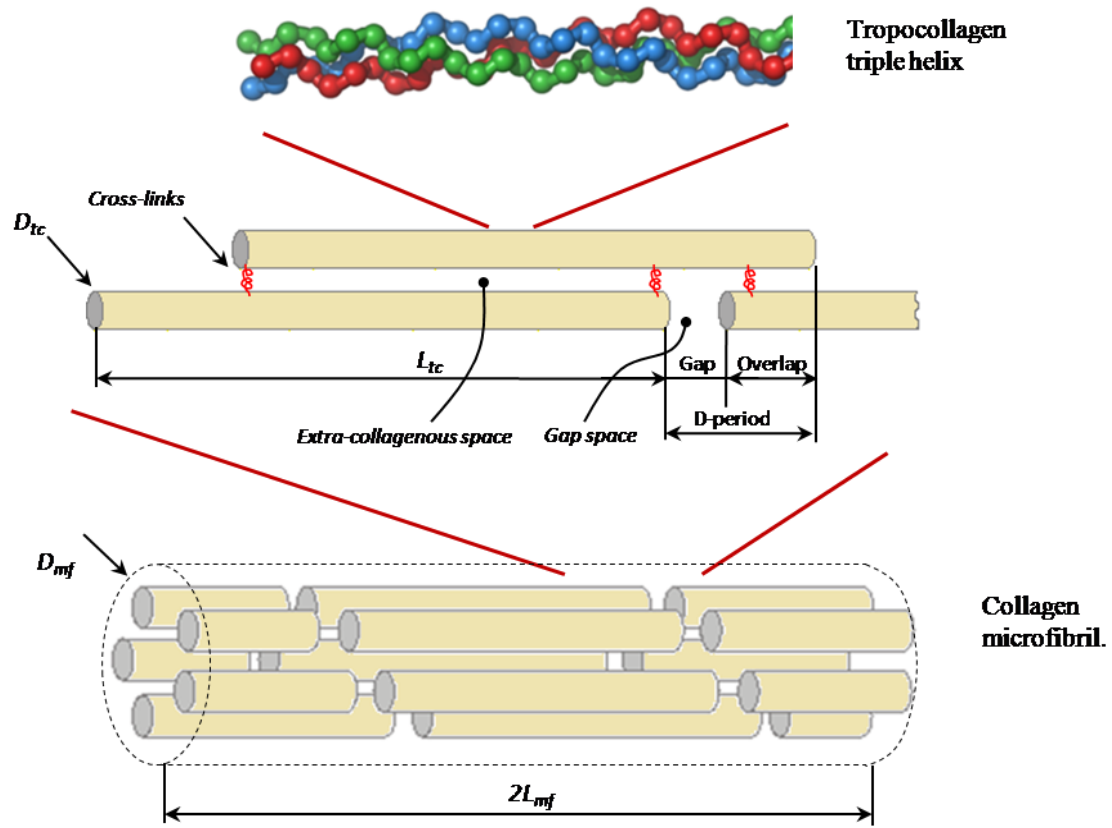

Figure 2. Double period model of cylindrical microfibril composed of (i) five TC molecules shifted by the interval D forming a cylindrical shape with, (ii) mineral phase filling the gap space (GM model) and the extra-collagenous space (EM model) and (iii) Cross-links joining two TC molecules ends.

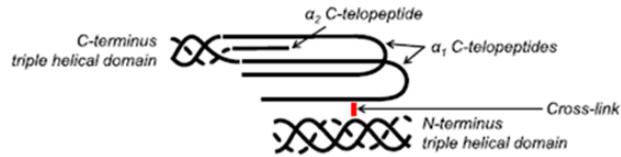
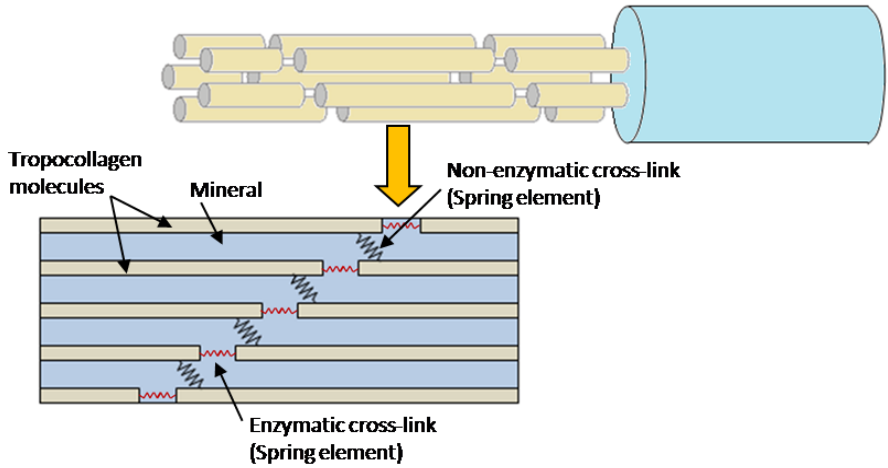

Figure 3. FE modeling of enzymatic and non-enzymatic cross links. Calibrated spring elements were used here based on experimental and theoretical data.

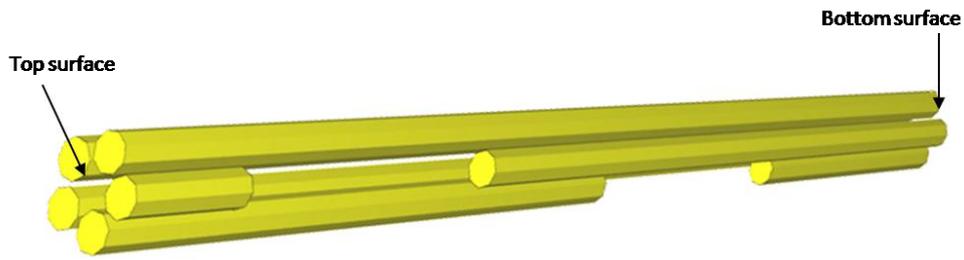

(a) **Model TC:** Microfibril 3D FE model composed of pure TC (without mineral) linked with crosslinks.

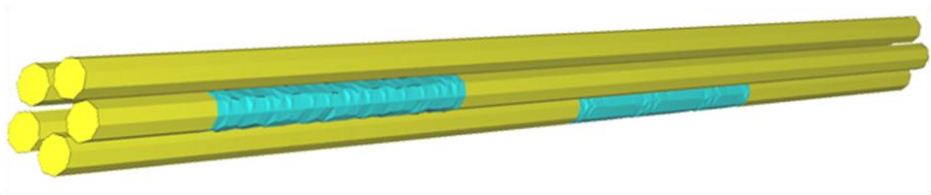

(b) **Model GM** (First stage mineral formation): Microfibril 3D FE model composed with TC, mineral filling the gap region of the TC and cross-links.

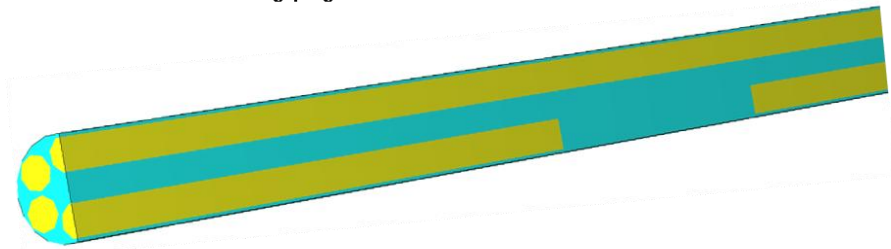

(c) **Model EM** (second stage mineral formation): Microfibril 3D FE model composed with TC, mineral filling the extra-collagenous space and cross-links.

Figure 4. Fraction of 3D FE periodic model of the microfibril composed of 35000 tetraedric elements. The FE mesh (small size of the tetraedric elements) and the spring elements (cross-links) are not plotted here for clarity. Due to the excessive microfibrile length to diameter ratio (340/4), only a segment of microfibril full length is plotted.

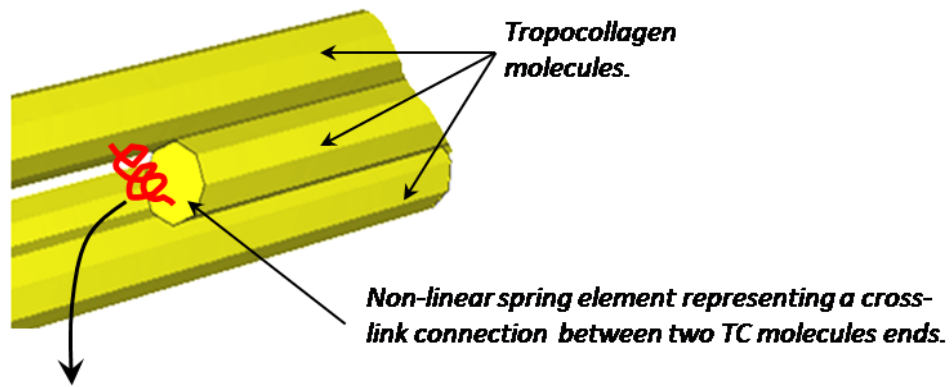

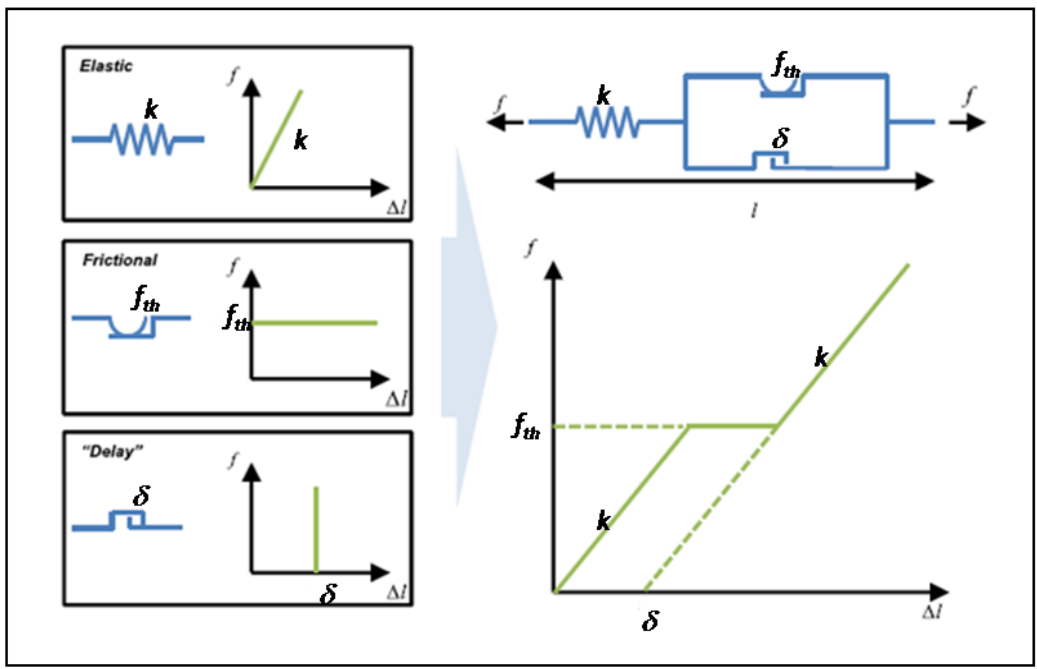

Figure 5. Rheological model of the non-linear spring element representing a cross-link behaviour with three regimes (i) elastic behaviour, (ii) delayed spring response and (iii) friction due to the intermolecular slippage from [12].

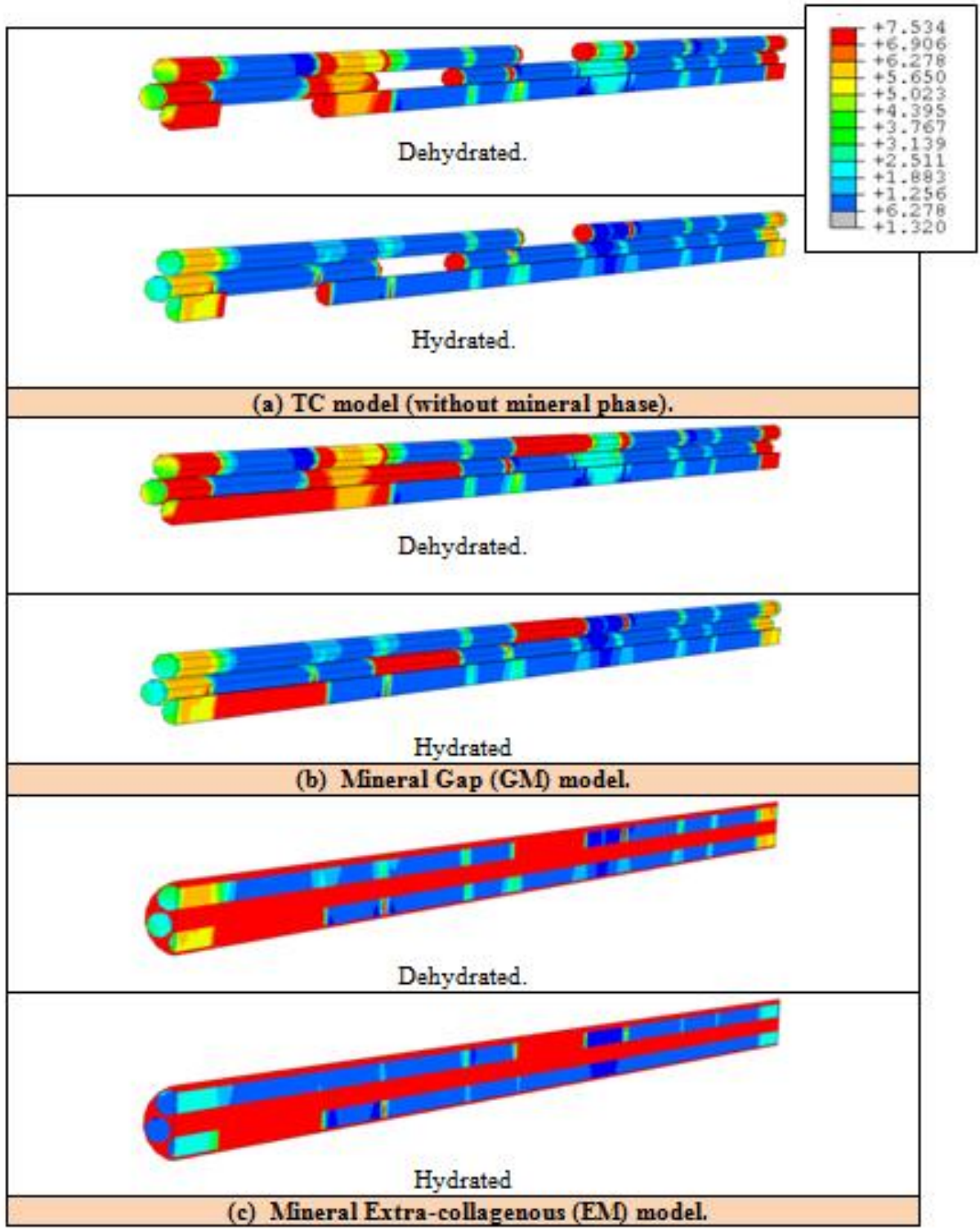

Figure 6. Contour of predicted equivalent von Mises stress (MPa) applied to the microfibril (Longitudinal cross section presentation) for three different models with hydrated and dehydrated TC molecules.

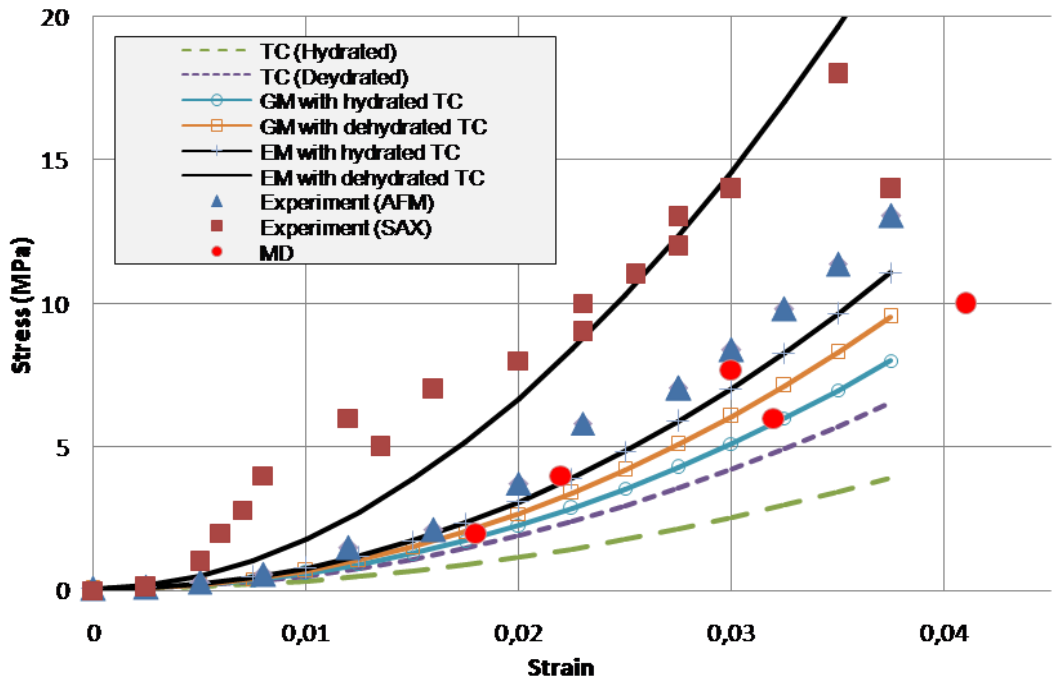

Figure 7. Microfibril tensile stress-strain behaviour at small strain regime for different models (dry and wet TC). Comparison between predicted 3D FE and experimental results (SAX) from [40], (AFM) from [25, 41] and molecular dynamics (MD) compuation from [11].

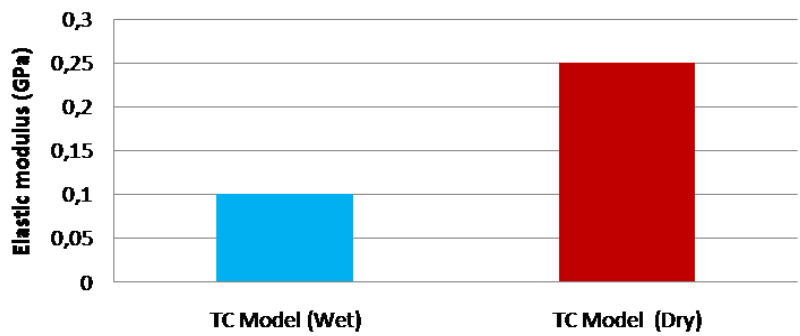

(a) Pure tropocollagen microfibril without mineral phase (TC model) for hydrated and dehydrated states.

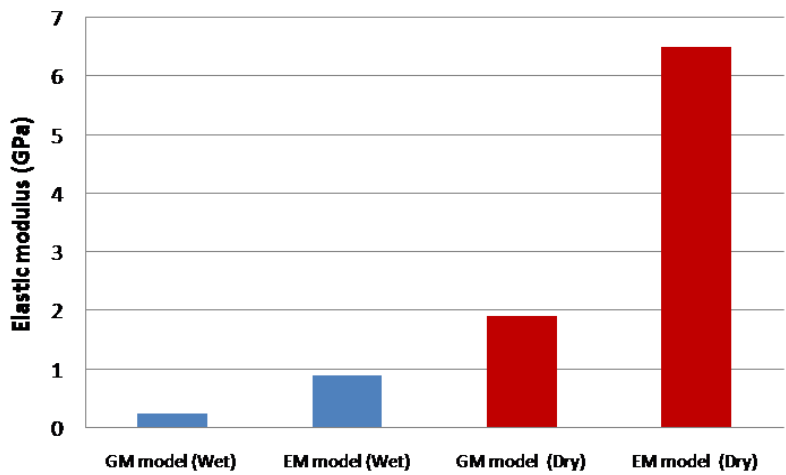

(b) Microfibril with mineral phases (GM and EM models) with TC (hydrated and dehydrated states).

Figure 8. Predicted elastic modulus for different microfibril models for wet and dry TC states. Increase of the mineral phase content within the microfibril combined with the TC state (dry or wet) leads to increase of the apparent modulus of the microfibril.

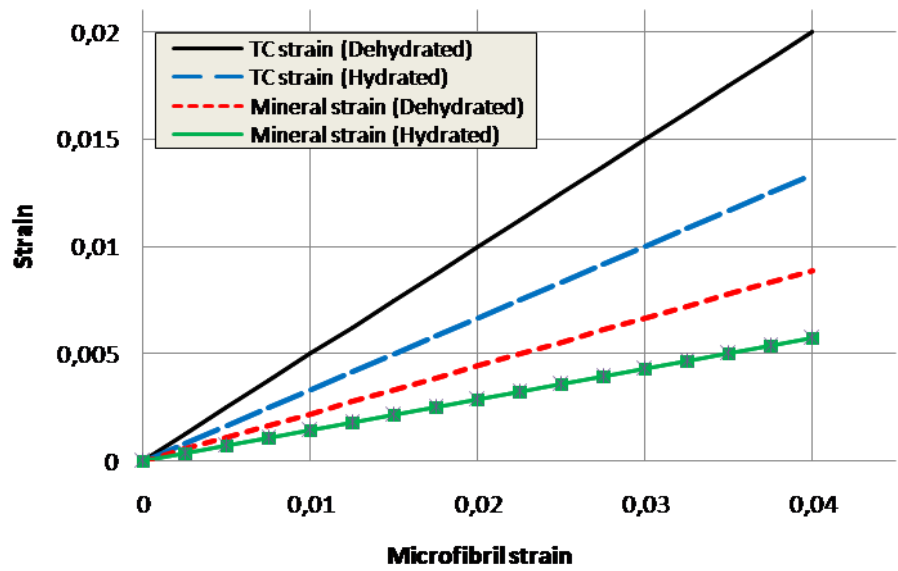

Figure 9. Correlation between TC molecule (dry and wet), mineral (GM model) strains and apparent microfibril strain.

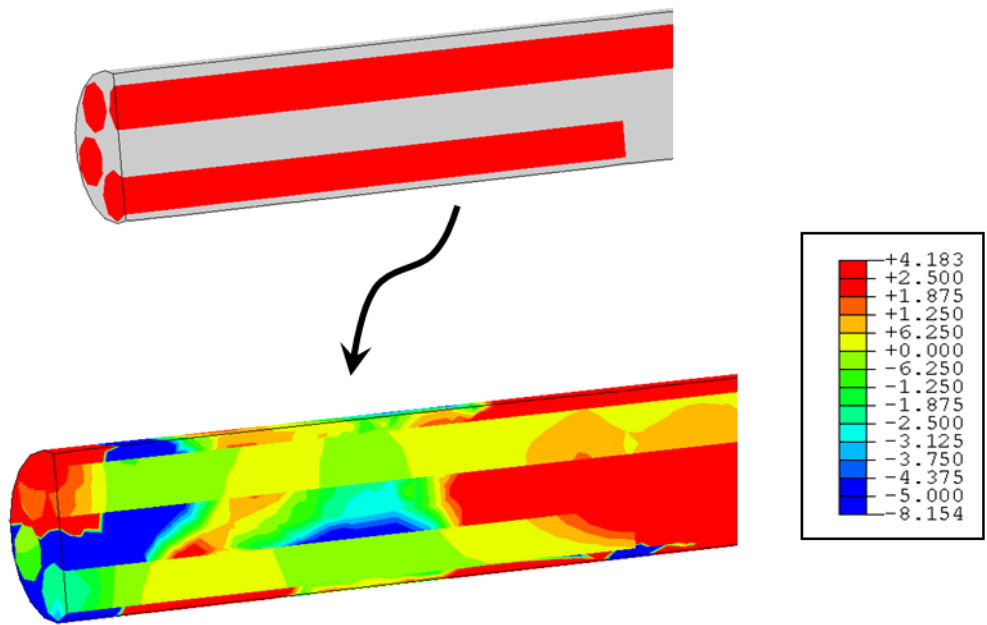

Figure 10. Longitudinal shear stress (in MPa) distribution in the top area of the microfibril (Zoom detail) showing the load transfer between the TC and the mineral.

| Description | Variable | Value (nm) | Source |
|---|---|---|---|
| **Length of model** | $L_{mf}$ | 340 | Buehler (2008), Gautieri et al. (2011)[10, 11] |
| **Diameter of model** | $D_{mf}$ | 4 | Buehler (2008), Gautieri et al. (2011) [10, 11] |
| **Length of tropocollagen molecule** | $L_{tc}$ | 300 | Buehler (2008), Gautieri et al. (2011) [10, 11] |
| **Diameter of tropocollagen molecule** | $D_{tc}$ | 1.09 (Dry) 1.5 (Wet) | Lees (1981) and Fratzl et al. (1993) [7,8] |
| **Periodicity** | D | 67 | Buehler (2008), Gautieri et al. (2011) [10, 11] |
| **Discrete gaps (hole zone)** | G (≈ 0.66 D) | 44.22 | Buehler (2008), Gautieri et al. (2011) [10, 11] |
| **Overlap** | O (≈ 0.34 D) | 22.78 | Buehler (2008), Gautieri et al. (2011) [10, 11] |

Table 1. Microfibril geometrical model variables and dimensions and their sources.

| Mechanical properties | Young's modulus (GPa) | Poisson's Ratio | Stress at fracture | Source |
|---|---|---|---|---|
| **Tropocollagen** | Small deformation: 2.8 | 0.30 | 9.3 GPa | Sasaki and Odajima (1996b) [74] |
| **Mineral** | 114 | 0.30 | 100 MPa | Weiner and Wagner (1998) [45] |

Table 2. Tropocollagen and mineral mechanical properties

| Rheological properties | Value | Source |
|---|---|---|
| Stiffness k ( N/nm) | 1181.1 3 e-11 | Buehler (2008) [10] |
| Friction parameter $f_{th}$ (pN) | 466 | Buehler et al. (2008)[76] |
| Delayed spring response parameter $\delta$ (nm) | 10 | Uzel and Buehler, (2011) [12] |

Table 3. Spring element rheological parameters and their sources representing the cross link behaviour for the proposed FE model.